\documentclass[twocolumn,prd]{revtex4}

\usepackage{amsmath}
\usepackage{graphicx}

\begin{document}

\title{Neutron star equilibrium configurations within a fully relativistic theory with strong, weak, electromagnetic, and gravitational interactions}

\author{Riccardo Belvedere}
\affiliation{Dipartimento di Fisica and ICRA, Sapienza Universit\`a di Roma, P.le Aldo Moro 5, I--00185 Rome, Italy}
\affiliation{ICRANet, P.zza della Repubblica 10, I--65122 Pescara, Italy}

\author{Daniela Pugliese}
\affiliation{Dipartimento di Fisica and ICRA, Sapienza Universit\`a di Roma, P.le Aldo Moro 5, I--00185 Rome, Italy}
\affiliation{School of Mathematical Sciences, Queen Mary, University of London, Mile End Road, London E1 4NS, United Kingdom}

\author{Jorge A. Rueda}
\affiliation{Dipartimento di Fisica and ICRA, Sapienza Universit\`a di Roma, P.le Aldo Moro 5, I--00185 Rome, Italy}
\affiliation{ICRANet, P.zza della Repubblica 10, I--65122 Pescara, Italy}

\author{Remo Ruffini}
\email{ruffini@icra.it}
\affiliation{Dipartimento di Fisica and ICRA, Sapienza Universit\`a di Roma, P.le Aldo Moro 5, I--00185 Rome, Italy}
\affiliation{ICRANet, P.zza della Repubblica 10, I--65122 Pescara, Italy}

\author{She-Sheng Xue}
\affiliation{Dipartimento di Fisica and ICRA, Sapienza Universit\`a di Roma, P.le Aldo Moro 5, I--00185 Rome, Italy}
\affiliation{ICRANet, P.zza della Repubblica 10, I--65122 Pescara, Italy}

\date{\today}

\begin{abstract}
We formulate the equations of equilibrium of neutron stars taking into account strong, weak, electromagnetic, and gravitational interactions within the framework of general relativity. The nuclear interactions are described by the exchange of the $\sigma$, $\omega$, and $\rho$ virtual mesons. The equilibrium conditions are given by our recently developed theoretical framework based on the Einstein-Maxwell-Thomas-Fermi equations along with the constancy of the general relativistic Fermi energies of particles, the ``Klein potentials'', throughout the configuration. The equations are solved numerically in the case of zero temperatures and for selected parameterizations of the nuclear models. The solutions lead to a new structure of the star: a positively charged core at supranuclear densities surrounded by an electronic distribution of thickness $\sim \hbar/(m_e c)\sim 10^2\hbar/(m_\pi c)$ of opposite charge, as well as a neutral crust at lower densities. Inside the core there is a Coulomb potential well of depth $\sim m_\pi c^2/e$. The constancy of the Klein potentials in the transition from the core to the crust, impose the presence of an overcritical electric field $\sim (m_\pi/m_e)^2 E_c$, the critical field being $E_c=m^2_e c^3/(e \hbar)$. The electron chemical potential and the density decrease, in the boundary interface, until values $\mu^{\rm crust}_e < \mu^{\rm core}_e$ and $\rho_{\rm crust}<\rho_{\rm core}$. For each central density, an entire family of core-crust interface boundaries and, correspondingly, an entire family of crusts with different mass and thickness, exist. The configuration with $\rho_{\rm crust}=\rho_{\rm drip}\sim 4.3\times 10^{11}$ g/cm$^3$ separates neutron stars with and without inner crust. We present here the novel neutron star mass-radius for the especial case $\rho_{\rm crust}=\rho_{\rm drip}$ and compare and contrast it with the one obtained from the traditional Tolman-Oppenheimer-Volkoff treatment.
\end{abstract}

\maketitle

\section{Introduction}\label{sec:1}	
%
It is well known that the classic works of Tolman \cite{tolman39} and of Oppenheimer and Volkoff \cite{oppenheimer39}, for short TOV, addresses the problem of neutron star equilibrium configurations composed only of neutrons. For the more general case when protons and electrons are also considered, in all of the scientific literature on neutron stars it is assumed that the condition of local charge neutrality applies identically to all points of the equilibrium configuration (see e.g.~\cite{nsbook}). Consequently, the corresponding solutions in this more general case of a non-rotating neutron star, are systematically obtained also on the base of the TOV equations. 

In general, the formulation of the equilibrium of systems composed by different particle species must be established within the framework of statistical physics of multicomponent systems; see e.g.~\cite{evans92}. Thermodynamic equilibrium of these systems is warrantied by demanding the constancy throughout the configuration of the generalized chemical potentials, often called ``electro-chemical'', of each of the components of the system; see e.g.~\cite{klein49,kodama72,olson75}. Such generalized potentials include not only the contribution due to kinetic energy but also the contribution due to the potential fields, e.g. gravitational and electromagnetic potential energies per particle, and in the case of rotating stars also the centrifugal potential. For such systems in presence of gravitational and Coulomb fields, global electric polarization effects at macroscopic scales occur. The balance of the gravitational and electric forces acting on ions and electrons in ideal electron-ion plasma leading to the occurrence of gravito-polarization was pointed out in the classic work of S.~Rosseland \cite{rosseland24}. 

If one turns to consider the gravito-polarization effects in neutron stars, the corresponding theoretical treatment acquires remarkable conceptual and theoretical complexity, since it must be necessarily formulated consistently within the Einstein-Maxwell system of equations. O.~Klein, in \cite{klein49}, first introduced the constancy of the general relativistic chemical potential of particles, hereafter ``Klein potentials'', in the study of the thermodynamic equilibrium of a self-gravitating one-component fluid of neutral particles throughout the configuration within the framework of general relativity. The extension of the Klein's work to the case of neutral multicomponent degenerate fluids can be found in \cite{{kodama72}} and to the case of multi-component degenerate fluid of charged particles in \cite{olson75}.

Using the concept of Klein potentials, we have recently proved the impossibility of imposing the condition of local charge neutrality in the simplest case of a self-gravitating system of degenerate neutrons, protons and electrons in $\beta$-equilibrium \cite{PLB2011}: it has been shown that the consistent treatment of the above system implies the solution of the general relativistic Thomas-Fermi equations, coupled with the Einstein-Maxwell ones, being the TOV equations thus superseded. 

We have recently formulated the theory of a system of neutrons, protons and electrons fulfilling strong, electromagnetic, weak and gravitational interactions \cite{NPA2011}. The role of the Klein first integrals has been again evidenced and their theoretical formulation in the Einstein-Maxwell background and in the most general case of finite temperature has been there presented, generalizing the previous results for the ``non-interacting'' case \cite{PLB2011}. The strong interactions, modeled by a relativistic nuclear theory, are there described by the introduction of the $\sigma$, $\omega$ and $\rho$ virtual mesons \cite{duerr56,walecka74,bowers73a,bowers73b} (see Subsec.~\ref{subsec:core} for details).

In this article we construct for the first time the equilibrium configurations of non-rotating neutron stars following the new approach, \cite{PLB2011,NPA2011}. The full set of the Einstein-Maxwell-Thomas-Fermi equations is solved numerically for zero temperatures and for selected parameterizations of the nuclear model. We use units with $\hbar=c=1$ throughout the article.

\section{The Constitutive Relativistic Equations}\label{sec:2}

\subsection{Core Equations}\label{subsec:core}

It has been clearly recognized that, since neutron stars cores may reach density of order $\sim 10^{16}$--$10^{17}$ g/cm$^3$, much larger than the nuclear density $\rho_{\rm nuc} \sim 2.7 \times 10^{14}$ g/cm$^3$, approaches for the nuclear interaction between nucleons based on phenomenological potentials and non-relativistic many-body theories become inapplicable (see \cite{bowers73a,bowers73b}). A self-consistent relativistic and well-tested model for the nuclear interactions has been formulated in \cite{duerr56,walecka74,bowers73a,bowers73b}. Within this model the nucleons interact with $\sigma$, $\omega$ and $\rho$ mesons through Yukawa-like couplings and assuming flat spacetime the equation of state of nuclear matter has been determined. However, it has been clearly stated in \cite{PLB2011,NPA2011} that, when we turn into a neutron star configuration at nuclear and supranuclear, the global description of the Einstein-Maxwell-Thomas-Fermi equations is mandatory. Associated to this system of equations there is a sophisticated eigenvalue problem, especially the one for the general relativistic Thomas-Fermi equation is necessary in order to fulfill the global charge neutrality of the system and to consistently describe the confinement of the ultrarelativistic electrons. 

The strong interactions between nucleons are described by the exchange of three virtual mesons: $\sigma$ is an isoscalar meson field providing the attractive long-range part of the nuclear force; $\omega$ is a massive vector field that models the repulsive short range and; $\rho$ is the massive isovector field that takes account surface as well as isospin effects of nuclei (see also \cite{boguta77,ring96}).

The total Lagrangian density of the system is given by 
\begin{equation}\label{Eq:TLD}
\mathcal{L}=\mathcal{L}_{g}+\mathcal{L}_{f}+\mathcal{L}_{\sigma}+\mathcal{L}_{\omega}+\mathcal{L}_{\rho}+\mathcal{L}_{\gamma}+\mathcal{L}_{\rm int},
\end{equation}
where the Lagrangian densities for the free-fields are
\begin{eqnarray}
\mathcal{L}_g &=& -\frac{R}{16 \pi G},  \label{eq:Lg}\\
\mathcal{L}_{\gamma} &=& -\frac{1}{16\pi} F_{\mu\nu}F^{\mu\nu},  \label{eq:Lgamma}\\
\mathcal{L}_{\sigma} &=& \frac{1}{2} \nabla_{\mu}\sigma \nabla^{\mu}\sigma-U(\sigma),  \label{eq:Ls}\\
\mathcal{L}_{\omega} &=& -\frac{1}{4} \Omega_{\mu\nu}\Omega^{\mu\nu}+\frac{1}{2}m_{\omega}^{2} \omega_{\mu} \omega^{\mu},  \label{eq:Lomega} \\
\mathcal{L}_{\rho} &=& -\frac{1}{4} \mathcal{R}_{\mu\nu}\mathcal{R}^{\mu\nu}+\frac{1}{2}m_{\rho}^{2} \rho_{\mu} \rho^{\mu},   \label{eq:Lrho}
\end{eqnarray}
where $\Omega_{\mu\nu}\equiv\partial_{\mu}\omega_{\nu}-\partial_{\nu}\omega_{\mu}$, $\mathcal{R}_{\mu\nu}\equiv\partial_{\mu}\rho_{\nu}-\partial_{\nu}\rho_{\mu}$,
$F_{\mu\nu}\equiv\partial_{\mu}A_{\nu}-\partial_{\nu}A_{\mu}$ are the field strength tensors for the
$\omega^{\mu}$, $\rho$ and $A^{\mu}$ fields respectively, $\nabla_\mu$ stands for covariant derivative and $R$ is the Ricci scalar. We adopt the Lorenz gauge for the fields $A_\mu$, $\omega_\mu$, and $\rho_\mu$. The self-interaction scalar field potential $U(\sigma)$ is a quartic-order polynom for a renormalizable theory (see e.g.~\cite{lee74}).

The Lagrangian density for the three fermion species is
\begin{equation}
\mathcal{L}_f = \sum_{i=e, N}\bar{\psi}_{i}\left(i \gamma^\mu D_\mu-m_i \right)\psi_i, \label{eq:Lf}
\end{equation}
where $\psi_N$ is the nucleon isospin doublet, $\psi_e$ is the electronic singlet, $m_i$ states for the mass of each particle-specie and $D_\mu = \partial_\mu + \Gamma_\mu$, being $\Gamma_\mu$ the Dirac spin connections.

The interacting part of the Lagrangian density is, in the minimal coupling assumption, given by
\begin{eqnarray}\label{eq:Lint}
\mathcal{L}_{\rm int} &=& -g_{\sigma} \sigma \bar{\psi}_N \psi_N - g_{\omega} \omega_{\mu} J_{\omega}^{\mu}-g_{\rho}\rho_{\mu}J_{\rho}^{\mu} \nonumber\\
&+& e A_{\mu} J_{\gamma,e}^{\mu}-e A_{\mu} J_{\gamma,N}^{\mu},
\end{eqnarray}
where the conserved currents are
\begin{eqnarray}
J^{\mu}_{\omega} &=& \bar{\psi}_N \gamma^{\mu}\psi_N, \label{eq:J1}\\
J^{\mu}_{\rho} &=& \bar{\psi}_N \tau_3\gamma^{\mu}\psi_N, \label{eq:J2}\\
J^{\mu}_{\gamma, e} &=& \bar{\psi}_e \gamma^{\mu}\psi_e, \label{eq:J3}\\
J^{\mu}_{\gamma, N} &=& \bar{\psi}_N \left(\frac{1+\tau_3}{2}\right)\gamma^{\mu}\psi_N. \label{eq:J4}
\end{eqnarray}

The coupling constants of the $\sigma$, $\omega$ and $\rho$-fields are $g_{\sigma}$, $g_{\omega}$ and $g_{\rho}$, and $e$ is the fundamental electric charge. The  Dirac matrices $\gamma^{\mu}$ and the isospin Pauli matrices satisfy the Dirac algebra in curved spacetime (see e.g.~\cite{lee87} for details).

We first introduce the non-rotating spherically symmetric spacetime metric
\begin{equation}\label{Eq:metriSN}
ds^2= {\rm e}^{\nu(r)}dt^2-{\rm e}^{\lambda(r)}dr^2-r^2d\theta^2 - r^2 \sin^2\theta d\varphi^2,
\end{equation}
where the $\nu(r)$ and $\lambda(r)$ are only functions of the radial coordinate $r$.

For very large number of fermions, we adopt the mean-field approximation in which fermion-field operators are replaced by their expectation values (see \cite{ruffini69} for details). Within this approximation, the full system of general relativistic equations can be written in the form
\begin{eqnarray}
&&{\rm e}^{-\lambda(r)}\left(\frac{1}{r^2}-\frac{1}{r}\frac{d\lambda}{dr}\right)-\frac{1}{r^2}= -8 \pi G T_0^0, \label{eq:EMTF1}\\
&&{\rm e}^{-\lambda(r)}\left(\frac{1}{r^2}+\frac{1}{r}\frac{d\nu}{dr}\right)-\frac{1}{r^2}= -8 \pi G T_1^1, \label{eq:EMTF2}\\
&&V'' + \frac{2}{r}V' \left[ 1 - \frac{r (\nu'+\lambda')}{4}\right] =\nonumber\\
&&- 4 \pi e \, {\rm e}^{\nu/2} {\rm e}^{\lambda} (n_p -n_e) ,\label{eq:GRTF} \\
&&\frac{d^2\sigma}{dr^2}+\frac{d\sigma}{dr}\left[\frac{2}{r}+\frac{1}{2}\left(\frac{d\nu}{dr}-\frac{d\lambda}{dr}\right)\right]=\nonumber\\
&&{\rm e}^{\lambda}\left[\partial_{\sigma}U(\sigma)+g_s n_s\right], \label{eq:EMTF3}\\
&&\frac{d^2\omega}{dr^2}+\frac{d\omega}{dr}\left[\frac{2}{r}-\frac{1}{2}\left(\frac{d\nu}{dr}+\frac{d\lambda}{dr}\right)\right]=\nonumber\\
&&-{\rm e}^{\lambda}\left(g_{\omega}J^{\omega}_0-m_{\omega}^2\omega\right), \label{eq:EMTF4}\\
&&\frac{d^2\rho}{dr^2}+\frac{d\rho}{dr}\left[\frac{2}{r}-\frac{1}{2}\left(\frac{d\nu}{dr}+\frac{d\lambda}{dr}\right)\right]=\nonumber\\
&&-{\rm e}^{\lambda}\left(g_{\rho} J^{\rho}_0-m_{\rho}^2\rho\right), \label{eq:EMTF5} \\
&&E^F_e={\rm e}^{\nu/2}\mu_e -eV={\rm constant},\label{eq:ef12}\\
&&E^F_p ={\rm e}^{\nu/2}\mu_{p} + \mathcal{V}_{p} = {\rm constant}, ,\label{eq:ef13}\\
&&E^F_n ={\rm e}^{\nu/2}\mu_{n} + \mathcal{V}_{n} = {\rm constant}, ,\label{eq:ef14}
\end{eqnarray}
where we have introduced the notation $\omega_0=\omega$, $\rho_0=\rho$, and $A_0=V$ for the temporal components of the meson-fields. Here $\mu_i= \partial\mathcal{E}/\partial n_i = \sqrt{(P_i^F)^2+\tilde{m}^2_i}$ and $n_i=(P_i^F)^3/(3 \pi^2)$ are the free-chemical potential and number density of the $i$-specie with Fermi momentum $P_i^F$. The particle effective mass is $\tilde{m}_N = m_N + g_s \sigma$ and $\tilde{m}_e=m_e$ and the effective potentials $\mathcal{V}_{p,n}$ are given by 
\begin{eqnarray}\label{eq:effpot}
\mathcal{V}_{p} &=& g_{\omega}\omega + g_{\rho} \rho + eV\, ,\\
\mathcal{V}_{n} &=& g_{\omega}\omega - g_{\rho} \rho\, .
\end{eqnarray}

The constancy of the generalized Fermi energies $E^F_n$, $E^F_p$ and $E^F_e$, the Klein potentials, derives from the thermodynamic equilibrium conditions given by the statistical physics of multicomponent systems, applied to a system of degenerate neutrons, protons, and electrons within the framework of general relativity (see \cite{NPA2011} for details). These constants are linked by the $\beta$-equilibrium between the matter constituents
\begin{equation}\label{eq:betaeq}
E^F_n=E^F_p+E^F_e\, .
\end{equation}

The electron density $n_e$ is, via Eq.~(\ref{eq:ef12}), given by 
\begin{equation}\label{eq:ne}
n_e = \frac{{\rm e}^{-3 \nu/2}}{3 \pi^2}[\hat{V}^2 + 2 m_e \hat{V} - m^2_e ({\rm e}^{\nu}-1)]^{3/2}\, ,
\end{equation}
where $\hat{V} \equiv e V + E^F_e$. Substituting Eq.(~\ref{eq:ne}) into Eq.~(\ref{eq:GRTF}) one obtains the general relativistic extension of the relativistic Thomas-Fermi equation recently introduced for the study of compressed atoms \cite{PRC2011,wd2011}. This system of equations has to be solved with the boundary condition of global neutrality; see \cite{PLB2011,NPA2011} and below for details.

The scalar density $n_s$, within the mean-field approximation, is given by the following expectation value
\begin{equation}\label{eq:ns}
n_s = \langle\bar{\psi}_N \psi_N\rangle = \frac{2}{(2 \pi)^3}\sum_{i=n,p}\int d^3k\frac{\tilde{m}_N}{\epsilon_i(p)},
\end{equation}
where $\epsilon_i(p)=\sqrt{p^2+\tilde{m}^2_i}$ is the single particle energy. 

In the static case, only the temporal components of the covariant currents survive, i.e. $\langle \bar{\psi}(x)\gamma^{i}\psi(x)\rangle=0$. Thus, by taking the expectation values of Eqs.~(\ref{eq:J1})--(\ref{eq:J4}), we obtain the non-vanishing components of the currents
\begin{eqnarray}
J^{ch}_0 &=& n_{ch} u_0=(n_p-n_e)u_0, \label{eq:J12}\\
J^{\omega}_0 &=& n_b u_0=(n_n+n_p)u_0, \label{eq:J22}\\
J^{\rho}_0 &=& n_3 u_0=(n_p-n_n)u_0, \label{eq:J32}
\end{eqnarray}
where $n_b=n_p+n_n$ is the baryon number density and $u_0 = \sqrt{g_{00}} = e^{\nu/2}$ is the covariant temporal component of the four-velocity of the fluid, which satisfies $u^\mu u_\mu=1$.

The metric function $\lambda$ is related to the mass $M(r)$ and the electric field $E(r) = -{\rm e}^{-(\nu+\lambda)/2} V'$ through
\begin{eqnarray}\label{eq:lambda}
{\rm e}^{-\lambda(r)} &=& 1 - \frac{2 G M(r)}{r} + G r^2 E^2(r) \nonumber \\
&=& 1 - \frac{2 G M(r)}{r} + \frac{G Q^2(r)}{r^2}\, ,
\end{eqnarray}
being $Q(r)$ the conserved charge, related to the electric field by $Q(r)=r^2 E(r)$.

The energy-momentum tensor of free-fields and free-fermions $T^{\mu\nu}$ of the system is
\begin{equation}\label{eq:Tab}
T^{\mu \nu} =  T_{f}^{\mu\nu} + T_{\gamma}^{\mu\nu} + T_{\sigma}^{\mu\nu} + T_{\omega}^{\mu\nu} + T_{\rho}^{\mu\nu},
\end{equation}
where
\begin{eqnarray}
T_{\gamma}^{\mu\nu} &=& \frac{1}{4\pi}\left(F_{\alpha}^{\mu}F^{\alpha \nu}+\frac{1}{4}g^{\mu\nu}F_{\alpha\beta}F^{\alpha\beta}\right),
\\
T_{\sigma}^{\mu\nu} &=& \nabla^{\mu}\sigma\nabla^{\nu}\sigma-g^{\mu\nu}\left[\frac{1}{2}\nabla_{\sigma}\sigma\nabla^{\sigma}\sigma-U(\sigma)\right],\\
T_{\omega}^{\mu\nu} &=& \Omega_{\alpha}^{\mu}\Omega^{\alpha\nu}+\frac{1}{4}g^{\mu\nu}\Omega_{\alpha\beta}\Omega^{\alpha\beta} \nonumber \\
&+& m_{\omega}^2\left(\omega^{\mu}\omega^{\nu}-\frac{1}{2}g^{\mu\nu}\omega_{\alpha}\omega^{\alpha}\right),\\
T_{\rho}^{\mu\nu} &=& \mathcal{R}_{\alpha}^{\mu}\mathcal{R}^{\alpha\nu}+\frac{1}{4}g^{\mu\nu}\mathcal{R}_{\alpha\beta}\mathcal{R}^{\alpha\beta} \nonumber \\ &+& m_{\rho}^2\left(\mathcal{R}^{\mu}\mathcal{R}^{\nu}-\frac{1}{2}g^{\mu\nu}\mathcal{R}_{\alpha}\omega^{\alpha}\right), \\
T_{f}^{\mu\nu} &=& (\mathcal{E}+\mathcal{P})u^{\mu}u^{\nu}-\mathcal{P}g^{\mu\nu},
\end{eqnarray}
where the energy-density $\mathcal{E}$ and the pressure $\mathcal{P}$ are given by
\begin{equation}\label{eq:EOS1}
\mathcal{E}=\sum_{i=n,p,e}\mathcal{E}_i, \qquad \mathcal{P}=\sum_{i=n,p,e}\mathcal{P}_i,
\end{equation}
being $\mathcal{E}_i$ and $\mathcal{P}_i$ the single fermion fluid contributions
\begin{eqnarray}
\mathcal{E}_i &=&\frac{2}{(2 \pi)^3} \int_0^{P^F_i} \epsilon_i(p)\,4 \pi p^2 dp, \label{eq:EOS2a}\\
\mathcal{P}_i&=&\frac{1}{3} \frac{2}{(2 \pi)^3} \int_0^{P^F_i} \frac{p^2}{\epsilon_i(p)}\,4 \pi p^2 dp. \label{eq:EOS2b}
\end{eqnarray}

It is worth to recall that the equation of state (\ref{eq:EOS1})--(\ref{eq:EOS2b}) satisfies the thermodynamic law
\begin{equation}\label{eq:thermolaw}
\mathcal{E}+\mathcal{P}=\sum_{i=n,p,e}n_i \mu_i.
\end{equation}

The parameters of the nuclear model, namely the coupling constants $g_s$, $g_\omega$ and $g_\rho$, and the meson masses $m_\sigma$, $m_\omega$ and $m_\rho$ are usually fixed by fitting experimental properties of nuclei, e.g. saturation density, binding energy per nucleon (or experimental masses), symmetry energy, surface energy, and nuclear incompressibility. In Table \ref{tab:models} we present selected fits of the nuclear parameters. In particular, we show the following parameter sets: NL3 \cite{lalazissis97}, NL-SH \cite{sharma93}, TM1 \cite{sugahara94}, and TM2 \cite{hirata95}.

\begin{table}[h]
\begin{ruledtabular}
\begin{tabular}{lrrrr}
& NL3 & NL-SH & TM1 & TM2 \\
\hline

$m_\sigma$ (MeV) & 508.194 & 526.059 & 511.198 & 526.443\\
$m_\omega$ (MeV) & 782.501 & 783.000 & 783.000 & 783.000\\
$m_\rho$ (MeV) & 763.000 & 763.000 & 770.000 & 770.000\\
$g_s$ & 10.2170 & 10.4440 & 10.0289 & 11.4694\\
$g_\omega$ & 12.8680 & 12.9450 & 12.6139 & 14.6377\\
$g_\rho$ & 4.4740 & 4.3830 & 4.6322 & 4.6783\\
$g_2$ (fm$^{-1}$) & -10.4310 & -6.9099 & -7.2325 & -4.4440\\
$g_3$ & -28.8850 & -15.8337 & 0.6183 & 4.6076\\
$c_3$ & 0.0000 & 0.0000 & 71.3075 & 84.5318
\end{tabular}
\end{ruledtabular}
\caption{Selected parameter sets of the $\sigma$-$\omega$-$\rho$ model.}\label{tab:models}
\end{table}

The constants $g_2$ and $g_3$ are the third and fourth order constants of the self-scalar interaction as given by the scalar self-interaction potential
\begin{equation}
U(\sigma) = \frac{1}{2} m^2_\sigma \sigma^2 + \frac{1}{3} g_2 \sigma^3 + \frac{1}{4} g_3 \sigma^4\, .
\end{equation}
The non-zero constant $c_3$ that appears in the TM1 and TM2 models corresponds to the self-coupling constant of the  non-linear vector self-coupling $\frac{1}{4} c_3 (\omega_\mu \omega^\mu)^2$. We have not include such a self-coupling vector interaction in the general formulation presented above. However, we show also here the results of the integration when such a self-interaction is taken into account and we refer to \cite{sugahara94,hirata95} for details about the motivations of including that contribution.

The numerical integration of the core equations can be started given a central density and the regularity conditions at the origin; see below Sec.~\ref{sec:3} for details. At nuclear density the phase-transition to the ``solid'' crust takes place. Thus, the radius of the core $R_{\rm core}$ is given by ${\cal E}(r=R_{\rm core})/c^2 = \rho_{\rm nuc}$. These equations must be solved with the boundary conditions given by the fulfillment of the condition of global charge neutrality and the continuity of the Klein potentials of particles between the core and the crust.

\subsection{Core-crust transition layer equations}

In the core-crust interface, the mean-field approximation for the meson-fields is not valid any longer and thus a full numerical integration of the meson-field equations of motion, taking into account all gradient terms, must be performed. We expect the core-crust transition boundary-layer to be a region with characteristic length scale of the order of the electron Compton wavelength $\sim \lambda_e = \hbar/(m_e c) \sim 100$ fm corresponding to the electron screening scale. Then, in the core-crust transition layer, the system of equations (\ref{eq:EMTF1})--(\ref{eq:ef14}) reduces to 
\begin{eqnarray}
&& V''+ \frac{2}{r} V' = -e^{\lambda_{\rm core}} e J_{ch}^0\, ,\label{eq:V3}\\
&& \sigma'' + \frac{2}{r} \sigma' = e^{\lambda_{\rm core}}\left[\partial_{\sigma}U(\sigma)+g_s n_s\right]\, ,\label{eq:sigma3}\\
&& \omega'' + \frac{2}{r} \omega' = -e^{\lambda_{\rm core}}\left[g_{\omega}J^{\omega}_0-m_{\omega}^2\omega\right]\, ,\label{eq:omega3}\\
&& \rho'' + \frac{2}{r} \rho' = -e^{\lambda_{\rm core}}\left[g_{\rho} J^{\rho}_0-m_{\rho}^2\rho\right]\, ,\label{eq:rho3} \\
&& e^{\nu_{\rm core}/2}\mu_e -eV = {\rm constant}\, ,\label{eq:electroneq3} \\
&& e^{\nu_{\rm core}/2}\mu_p + eV + g_\omega \omega + g_\rho \rho = {\rm constant}\, ,\label{eq:protoneq3} \\
&& \mu_n = \mu_p + \mu_e + 2\, g_\rho \rho e^{-\nu_{\rm core}/2}\, ,\label{eq:betaeq3}
\end{eqnarray}
due to the fact that the metric functions are essentially constant on the core-crust transition layer and thus we can take their values at the core-radius $e^{\nu_{\rm core}} \equiv e^{\nu(R_{\rm core})}$ and $e^{\lambda_{\rm core}} \equiv e^{\lambda(R_{\rm core})}$.

\begin{figure}[h]
\includegraphics[width=\columnwidth]{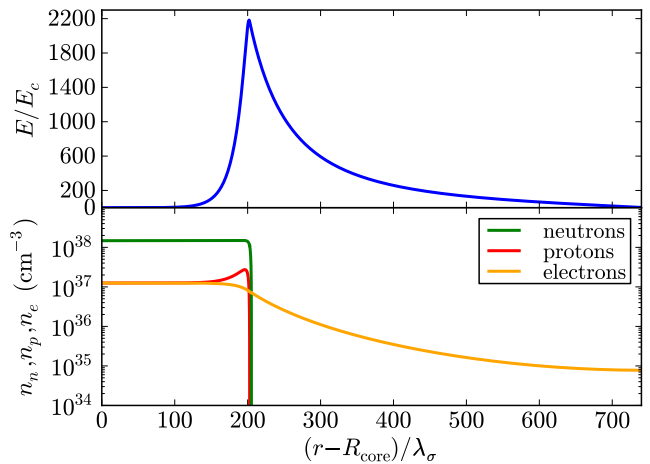}
\caption{Upper panel: electric field in the core-crust transition layer in units of the critical field $E_c$. Lower panel: particle density profiles in the core-crust boundary interface in units of cm$^{-3}$. Here we use the NL3-model of Table \ref{tab:models} and $\lambda_\sigma = \hbar/(m_\sigma c) \sim 0.4$ fm denotes the sigma-meson Compton wavelength. The density at the edge of the crust in this example is $\rho_{\rm crust}=\rho_{\rm drip}=4.3\times 10^{11}$ g/cm$^3$.}\label{fig:electric_field1}
\end{figure}

\begin{figure}[h]
\includegraphics[width=\columnwidth]{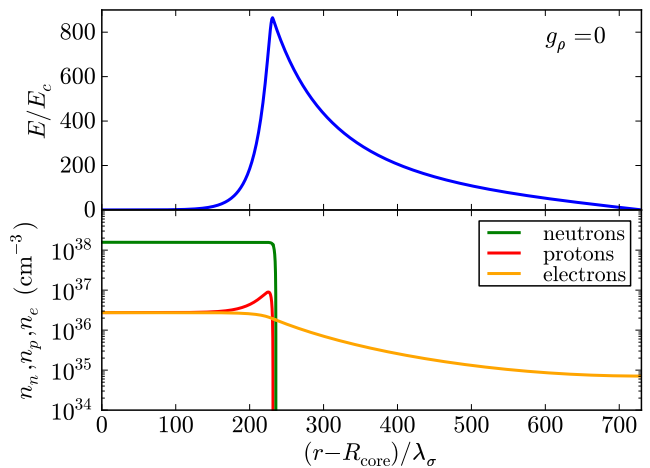}
\caption{The same as Fig.~\ref{fig:electric_field1}, but setting $g_\rho = 0$ in order to see the effects of the $\rho$-meson with respect to the case $g_\rho \neq 0$.}\label{fig:electric_field2}
\end{figure}

The system of equations of the transition layer has a stiff nature due to the existence of two different scale lengths. The first one is associated with the nuclear interactions $\sim \lambda_\pi = \hbar/(m_\pi c) \sim 1.5$ fm and the second one is due to the aforementioned screening length $\sim \lambda_e = \hbar/(m_e c) \sim 100$ fm. Thus, the numerical integration of Eqs.~(\ref{eq:V3})--(\ref{eq:betaeq3}) has been performed subdividing the core-crust transition layer in the following three regions: (I) a mean-field-like region where all the fields vary slowly with length scale $\sim \lambda_e$, (II) a strongly interacting region of scale $\sim \lambda_\pi$ where the surface tension due to nuclear interactions dominate producing a sudden decrease of the proton and the neutron densities and, (III) a Thomas-Fermi-like region of scale $\sim \lambda_e$ where only a layer of opposite charge made of electrons is present producing the total screening of the positively charged core. The results of the numerical integration of the equilibrium equations are shown in Fig.~\ref{fig:electric_field1}-\ref{fig:electric_field2} for the NL3-model.

We have integrated numerically Eqs.~(\ref{eq:EMTF1})--(\ref{eq:ef14}) for the models listed in Table \ref{tab:models}. The boundary conditions for the numerical integration are fixed through the following procedure. We start assuming a value for the central baryon number density $n_b(0) = n_n(0) + n_p(0)$. From the regularity conditions at the origin we have $e^{-\lambda(0)} = 1$ and $n_e(0) = n_p(0)$. 

The metric function $\nu$ at the origin can be chosen arbitrarily, e.g. $\nu(0)=0$, due to the fact that the system of equations remain invariant under the shift $\nu \to \nu +$ constant. The right value of $\nu$ is obtained once the end of the integration of the core has been accomplished and duly matched to the crust, by fulfilling the following identity at the surface of the neutron star,
\begin{equation}
e^{\nu(R)} =e^{-\lambda(R)} = 1-\frac{2 G M(R)}{c^2 R}\, ,
\end{equation}
being $M(R)$ and $R$ the total mass and radius of the star. Then, taking into account the above conditions, we solve the system (\ref{eq:EMTF3})--(\ref{eq:ef14}) at the origin for the other unknowns $\sigma(0)$, $\omega(0)$, $\rho(0)$, $n_n(0)$, $n_p(0)$, $n_e(0)$. 

The initial conditions for the numerical integration of the core-crust transition layer equations are determined by the final values given by the numerical integration of the core equations, i.e. we take the values of all the variables at the core-radius $R_{\rm core}$.

In the region I the effect of the Coulomb interaction is clear: on the proton-profile we can see a bump due to Coulomb repulsion while the electron-profile decreases as expected. Such a Coulomb effect is indirectly felt also by the neutrons due to the coupled nature of the system of equations. However, the neutron-bump is much smaller than the one of protons and it is not appreciable in Fig.~\ref{fig:electric_field1}-\ref{fig:electric_field2} due to the plot-scale. In the region II we see clearly the effect of the surface tension due to nuclear interaction which produces a sharp decrease of the neutron and proton profiles in a characteristic scale $\sim \lambda_\pi$. In addition, it can be seen a neutron skin effect, analogous to the one observed in heavy nuclei, which makes the scale of the neutron density falloff slightly larger with respect to the proton one, in close analogy to the neutron skin effect observed in neutron rich nuclei, see e.g.~\cite{PRLskin2011}. The region III is characterized by a smooth decreasing of the electron density which resembles the behavior of the electrons surrounding a nucleus in the Thomas-Fermi model.

The matching to the crust must be done at a radius $R_{\rm core} + \delta R$ where full charge neutrality of the core is reached. Different thicknesses $\delta R$ correspond to different electron Fermi energies $E^F_e$. The thickness of the core-crust transition boundary layer $\delta R$ as well as the value of the electron density at the edge of the crust, $n^{\rm crust}_e=n_e(R_{\rm core} + \delta R)$, depend on the nuclear parameters, especially on the nuclear surface tension. 

The equilibrium conditions given by the constancy of the Klein potentials (\ref{eq:ef12})--(\ref{eq:ef14}) throughout the configuration, impose in the transition layer the following continuity condition
\begin{equation}\label{eq:kleincontinuity}
E^F_e = e^{\nu_{\rm core}/2}\mu^{\rm core}_e - e V^{\rm core} = e^{\nu_{\rm crust}/2}\mu^{\rm crust}_e\, ,
\end{equation}
where $\mu^{\rm core}_e = \mu_e(R_{\rm core})$, $e V^{\rm core}=e V(R_{\rm core})$, and $\mu^{\rm crust}_e = \mu_e(R_{\rm core} + \delta R)$, and $e^{\nu_{\rm crust}} \simeq e^{\nu_{\rm core}}$. 

\begin{figure}[h]
\includegraphics[width=\columnwidth]{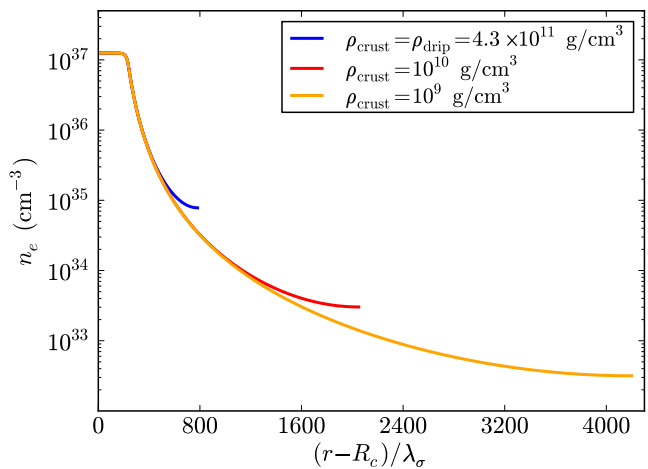}
\caption{Distribution of electrons in the core-crust boundary interface for different densities at the edge of the crust, $\rho_{\rm crust}$. The larger the $\rho_{\rm crust}$, the smaller the electric field $E$ and the smaller the thickness of the interface $\delta R$.}\label{fig:elecshell}
\end{figure}

In the boundary interface, the electron chemical potential and the density decrease: $\mu^{\rm crust}_e < \mu^{\rm core}_e$ and $\rho_{\rm crust}<\rho_{\rm core}$. For each central density, an entire family of core-crust interface boundaries exist each one with a specific value of $\delta R$: the larger the $\rho_{\rm crust}$, the smaller the $\delta R$. Correspondingly, an entire family of crusts with different mass and thickness, exist. From the continuity of the electron Klein potential in the boundary interface given by Eq.~(\ref{eq:kleincontinuity}), it follows that different values of $\rho_{\rm crust} \geq 0$ correspond to different values of the electron Fermi energy $E^F_e \geq 0$. In close analogy to the compressed atoms studied in \cite{PRC2011}, the case $E^F_e = 0$ corresponds to the ``free'' (uncompressed) configuration, where $\delta R \to \infty$ and $\rho_{\rm crust} = 0$, i.e. a bare core. In this configuration the electric field reaches its maximum value. The case $E_e^F>0$ is analogous to the one of the compressed atom \cite{PRC2011}. In Fig.~\ref{fig:elecshell} we have plotted the electron distribution in the core-crust boundary interface for selected densities at the edge of the crust $\rho_{\rm crust}=[\rho_{\rm drip},10^{10},10^9]$ g/cm$^3$, where $\rho_{\rm drip}\sim 4.3\times 10^{11}$ g/cm$^3$ is the neutron drip density.

The configuration with $\rho_{\rm crust}=\rho_{\rm drip}$ separates neutron stars with and without inner crust. In the so-called inner crust, the neutrons dripped from the nuclei in the crust form a fluid that coexist with the nuclei lattice and the degenerate electrons \cite{baym71}. 
For definiteness, we present in this article the results for configurations $\rho_{\rm crust}\leq \rho_{\rm drip}$, i.e for neutron stars possessing only outer crust. The construction of configurations with $\rho_{\rm crust}>\rho_{\rm drip}$ needs to be studied in more detail and will be the subject of a forthcoming work.

In Figs.~\ref{fig:electric_field1} and \ref{fig:electric_field2}, we show the core-crust transition layer for the NL3 model of Table \ref{tab:models} with and without the presence of the $\rho$-meson respectively. The presence of the $\rho$-meson is responsible for the nuclear asymmetry within this nuclear model. The relevance of the nuclear symmetry energy on the structure of nuclei and neutron stars is continuously stressed in literature; see e.g.~\cite{symmetry1,symmetry2,symmetry3,symmetry4,symmetry5}. The precise value of the nuclear symmetry energy plays here a crucial in determining the precise value of the $\rho$-meson coupling which, in the present case, is essential in the determination of the intensity of the electric field in the core-crust boundary interface; as can be seen from the comparison of Figs.~\ref{fig:electric_field1} and \ref{fig:electric_field2}.

\subsection{Crust equations}

Turning now to the crust, it is clear from our recent treatment of white dwarfs \cite{wd2011} that also this problem can be solved by the adoption of Wigner-Seitz cells and from the relativistic Feynman-Metropolis-Teller (RFMT) approach \cite{PRC2011} it follows that the crust is clearly neutral. Thus, the structure equations to be integrated are the TOV equations
\begin{eqnarray}
&& \frac{d\mathcal{P}}{dr}=-\frac{G(\mathcal{E}+\mathcal{P})(M+4\pi r^{3}\mathcal{P})}{r^2(1-\frac{2G M}{r})},\\
&& \frac{dM}{dr}=4\pi r^2 {\mathcal{E}},
\end{eqnarray}
where $M=M(r)$ is the mass enclosed at the radius $r$.

The effects of the Coulomb interaction in ``solid''-like electron-ion systems appears only at the microscopic level e.g.~Debye-Hueckel screening in classical systems \cite{debye23} and Thomas-Fermi screening in the degenerate case \cite{mott36}. In order to analyze the effects of the microscopic screening on the structure of the configuration we will consider two equations of state for the crust: the locally neutral case or uniform approximation (see e.g.~\cite{chandrasekhar31}) and, for simplicity, instead of using the RFMT EoS \cite{PRC2011}, we use as second EoS the one due to Baym, Pethick and Sutherland (BPS) \cite{baym71}, which is by far the most used equation of state in literature for the description of the neutron star crust (see e.g.~\cite{nsbook}).

In the uniform approximation, both the degenerate electrons and the nucleons distribution are considered constant inside each cell of volume $V_{\rm ws}$. This kind of configuration can be obtained only imposing microscopically the condition of local charge neutrality
\begin{equation}
n_e=\frac{Z}{V_{\rm ws}}.
\end{equation}

The total pressure of the system is assumed to be entirely due to the electrons, i.e.
\begin{equation}
\mathcal{P}=\mathcal{P}_e=\frac{2}{3\left(2\pi\hbar\right)^3}\int^{P_e^F}_{0}\frac{c^2p^24\pi p^2}{\sqrt{c^2p^2+m_e^2c^4}}dp,
\end{equation}
while the total energy-density of the system is due to the nuclei, i.e. $\mathcal{E}$=$(A/Z)m_N n_e$, where $m_N$ is the nucleon mass.

We turn now to the BPS equation of state. The first correction to the uniform model, corresponds to abandon the assumption of the electron-nucleon fluid through the so-called ``lattice'' model which introduces the concept of Wigner-Seitz cell: each cell of radius $R_{\rm ws}$ contains a point-like nucleus of charge $+Z e$ with $A$ nucleons surrounded by a uniformly distributed cloud of $Z$ fully-degenerate electrons.

The sequence of the equilibrium nuclides present at each density in the BPS equation of state is obtained by looking for the nuclear composition that minimizes the energy per nucleon for each fixed nuclear composition $(Z,A)$ (see Table \ref{tab:comparison} and \cite{baym71} for details). The pressure $\mathcal{P}$ and the energy-density ${\mathcal{E}}$ of the system are, within this model, given by 
\begin{eqnarray}
&& \mathcal{P}=\mathcal{P}_e + \frac{1}{3} W_Ln_N ,\\
&&\frac{\mathcal{E}}{n_b}=\frac{W_N+W_L}{A}+\frac{\mathcal{E}_e(n_bZ/A)}{n_b},
\end{eqnarray}
where the electron energy-density is given by
\begin{equation}
{\cal E}_e =  \frac{2}{(2 \pi)^3} \int_0^{P^F_e} \sqrt{p^2+m^2_e} 4 \pi p^2 dp, 
\end{equation}
and $W_N(A,Z)$ is the total energy of an isolated nucleus given by the semi-empirical formula
\begin{equation}
W_N=m_nc^2(A-Z)+m_pc^2Z-b A,
\end{equation}
with $b$ being the Myers and Swiatecki binding energy per nucleon \cite{myers66}. The lattice energy per nucleus $W_L$ is given by
\begin{equation}
W_L=-\frac{1.819620Z^2e^2}{a},
\end{equation}
where the lattice constant $a$ is related to the nucleon density $n_N$ by $n_N a^3=2$.

\section{Neutron star structure}\label{sec:3}

\begin{figure}[h]
\includegraphics[width=\columnwidth]{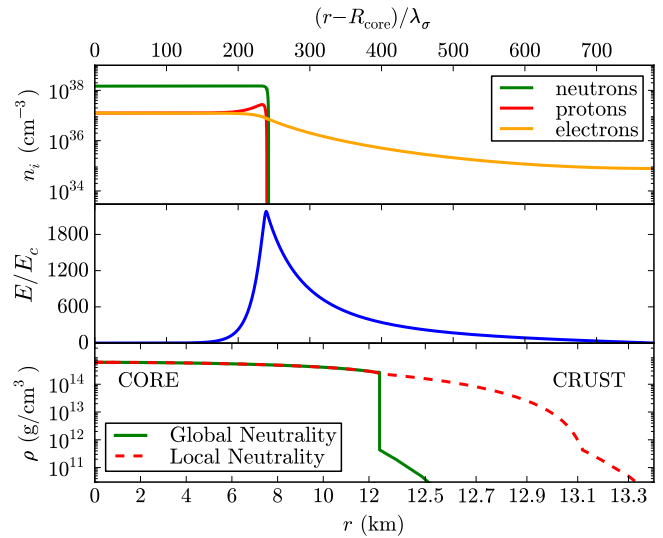}
\caption{Upper panel: electric field in the core-crust transition layer, in units of the critical field $E_c$. Middle panel: particle density profiles in the core-crust boundary interface, in units of cm$^{-3}$. Lower panel: density profile inside a neutron star with central density $\rho(0)\sim 5 \rho_{\rm nuc}$. We compare and contrast the structural differences between the solution obtained from the traditional TOV equations (locally neutral case) and the globally neutral solution presented here. We use here the NL3 nuclear parametrization of Table \ref{tab:models} and $\lambda_\sigma = \hbar/(m_\sigma c) \sim 0.4$ fm, denotes the sigma-meson Compton wavelength. In this example the density at the edge of the crust is $\rho_{\rm crust}=\rho_{\rm drip}=4.3\times 10^{11}$ g/cm$^3$.}\label{fig:density}
\end{figure}

\begin{figure}[h]
\includegraphics[width=\columnwidth]{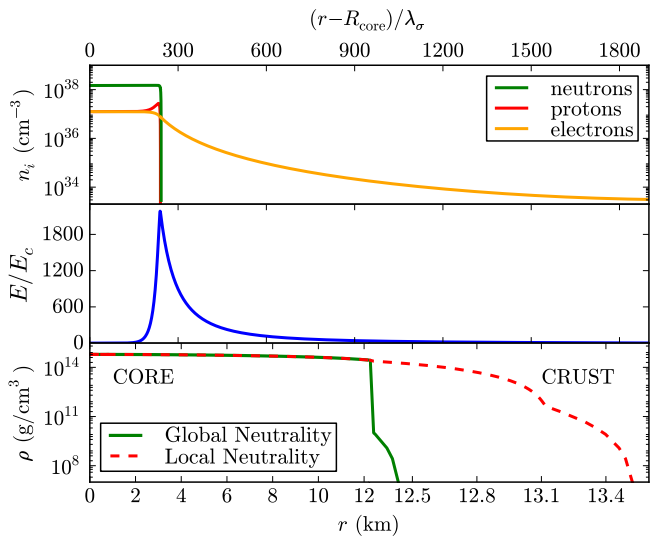}
\caption{Same as Fig.~\ref{fig:density}. In this example the density at the edge of the crust is $\rho_{\rm crust}=10^{10}$ g/cm$^3$.}\label{fig:density2}
\end{figure}

In the traditional TOV treatment the density and the pressure are a priori assumed to be continuous as well as the local charge neutrality of the system. The distinguishing feature of our new solution is that the Klein potentials are constant throughout the three regions; the core, the crust and the transition interface boundary. An overcritical electric field is formed and consequently a discontinuity in density is found with a continuous total pressure including the surface tension of the boundary. In Figs.~\ref{fig:density} and \ref{fig:density2}, we compare and contrast the density profiles of configurations obtained from the traditional TOV treatment and with the treatment presented here.

\begin{figure}[h]
\includegraphics[width=\columnwidth]{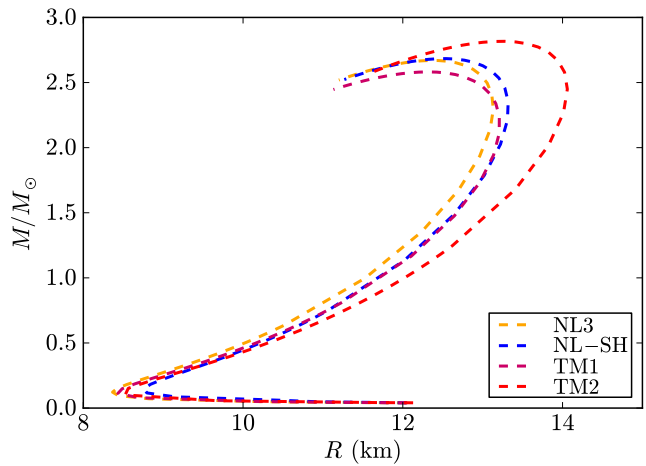}
\caption{Mass-Radius relation for the neutron stars obtained with the nuclear models listed in Table \ref{tab:models}. In the crust we have used the BPS equation of state. The mass is given in solar masses and the radius in km.}\label{fig:star}
\end{figure}

\begin{figure}[h]
\includegraphics[width=\columnwidth]{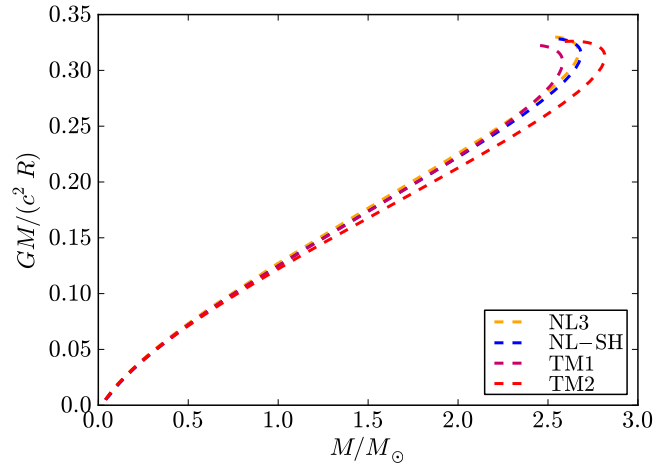}
\caption{Compactness of the star $G M/(c^2 R)$ as a function of the star mass $M$. In the crust we have used the BPS equation of state and the nuclear models are in Table \ref{tab:models}.\label{fig:compactness_star1}}
\end{figure}

\begin{figure}[h]
\includegraphics[width=\columnwidth]{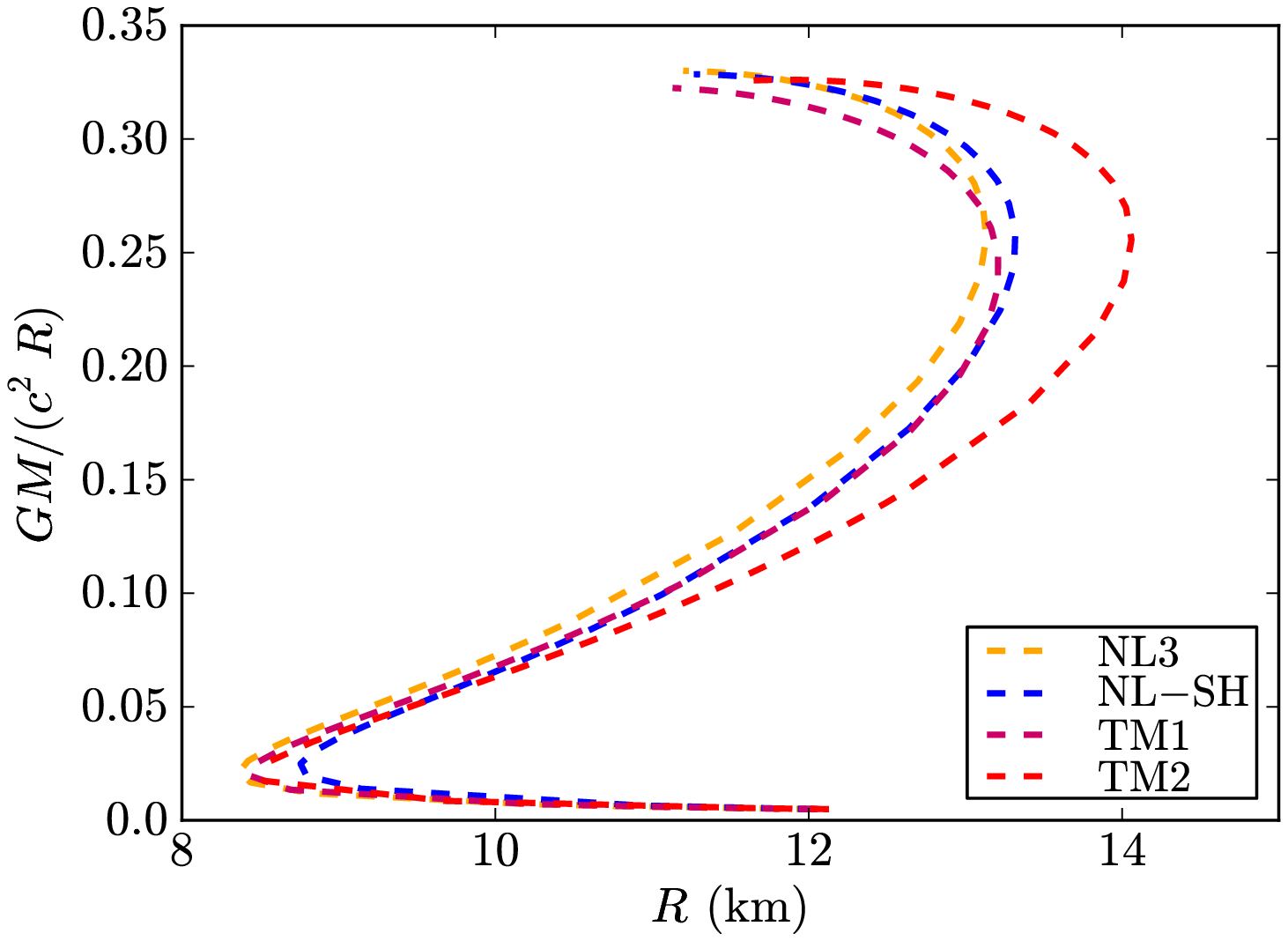}
\caption{Compactness of the star $G M/(c^2 R)$ as a function of the star radius $R$. In the crust we have used the BPS equation of state and the nuclear models are in Table \ref{tab:models}.\label{fig:compactness_star2}}
\end{figure}

\begin{figure}[h]
\includegraphics[width=\columnwidth]{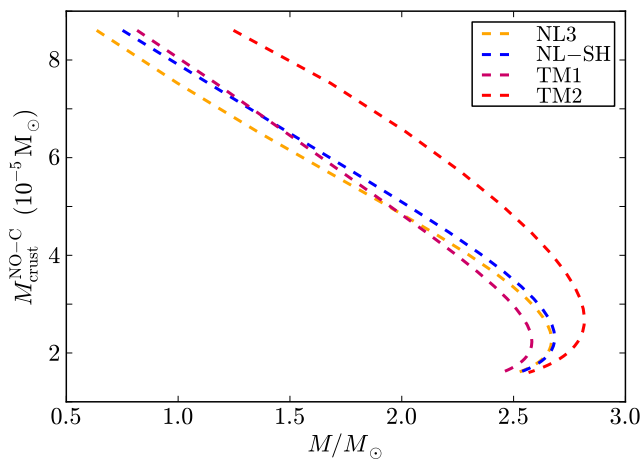}
\caption{Mass of the crust as a function of the compactness for the crust EoS without Coulomb interactions.}\label{fig:crust_NO-C1}
\end{figure}

\begin{figure}[h]
\includegraphics[width=\columnwidth]{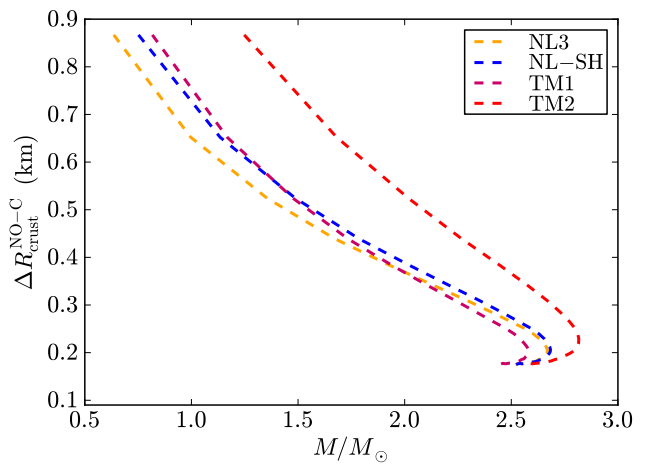}
\caption{Crust-thickness as a function of the compactness for the crust EoS without Coulomb interactions.}\label{fig:crust_NO-C2}
\end{figure}

\begin{figure}[h]
\includegraphics[width=\columnwidth]{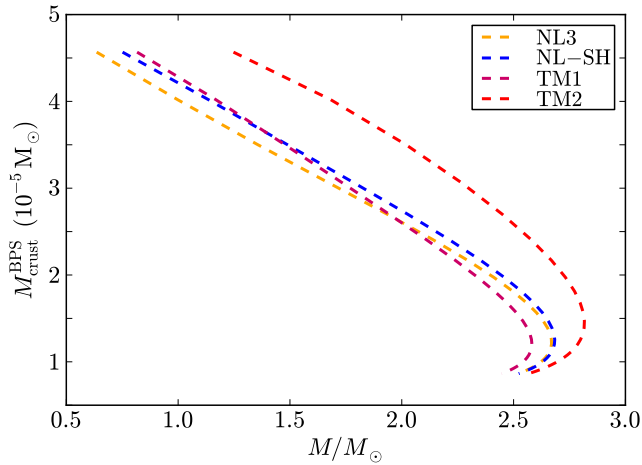}
\caption{Crust mass as a function of the compactness for crust with the BPS EoS.}\label{fig:crust_BPS1}
\end{figure}

\begin{figure}[h]
\includegraphics[width=\columnwidth]{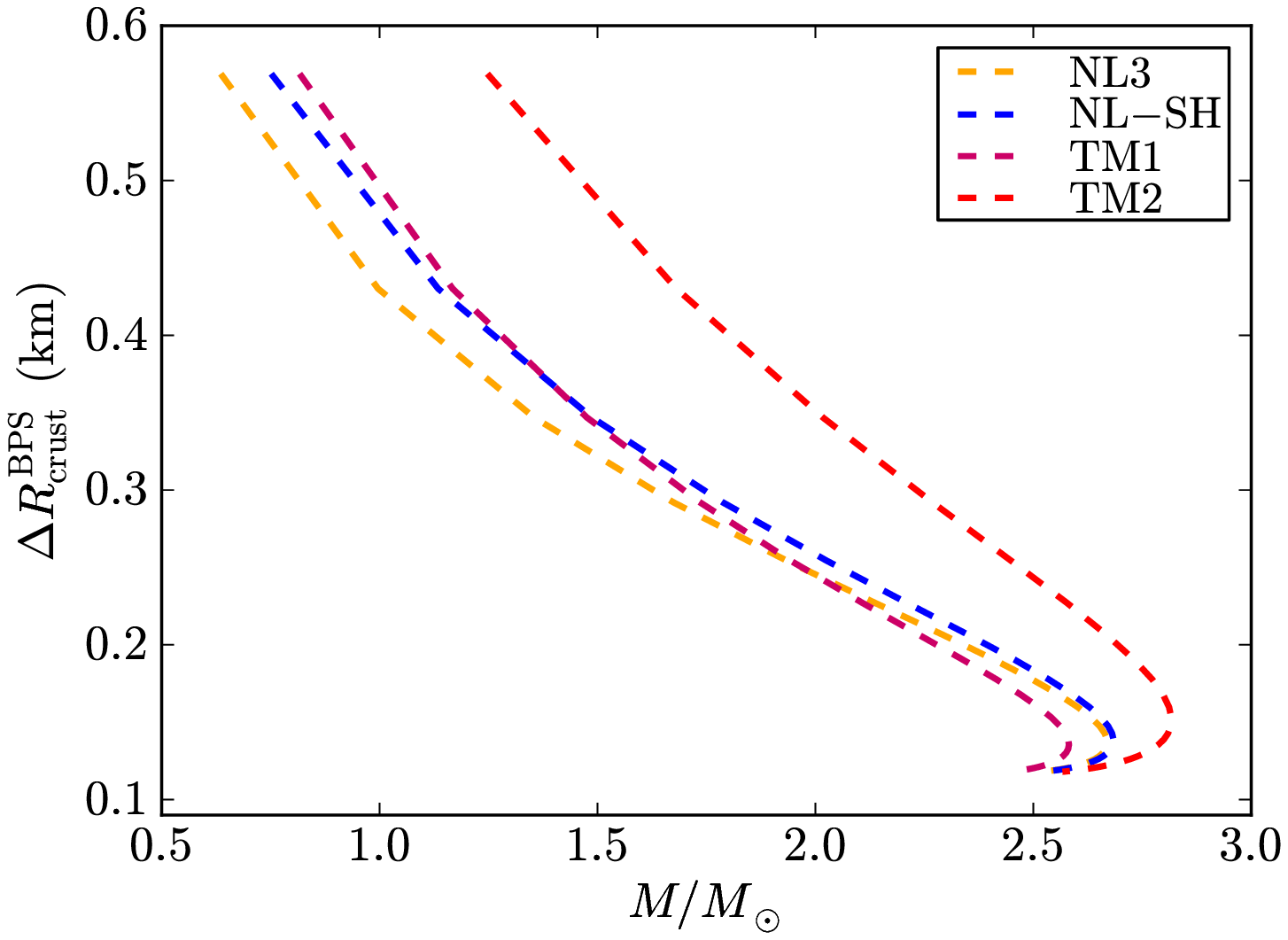}
\caption{Crust thickness as a function of the compactness for crust with the BPS EoS.}\label{fig:crust_BPS2}
\end{figure}

In Figs.~\ref{fig:star}--\ref{fig:crust_BPS2} we show the results of the numerical integration of the system of the general relativistic constitutive equations of the configuration from the center all the way up to the surface with the appropriate boundary conditions between the involved phases. In particular, we have plotted the mass-radius relation as well as the compactness of the neutron stars obtained with the models listed in Table \ref{tab:models}.

It is worth to note that the inclusion of the Coulomb interaction and in particular the presence of the negative lattice energy $W_L$ results in a decreasing of the pressure of the cells. Such an effect, as shown in Fig.~\ref{fig:crust_NO-C1}--\ref{fig:crust_BPS2}, leads to a decreasing of the mass and the thickness of the crust with respect to the uniform-approximation case where no Coulomb interactions are taken into account.

Comparing the mass and the thickness of the crust obtained with these two different EoS, we obtain systematically crusts with smaller mass and larger thickness when Coulomb interactions are taken into account. This results are in line with the recent results in \cite{wd2011}, where the mass-radius relation of white-dwarfs has been calculated using an EoS based on the relativistic Feynman-Metropolis-Teller model for compressed atoms \cite{PRC2011}. 

\begin{figure}[h]
\includegraphics[width=\columnwidth]{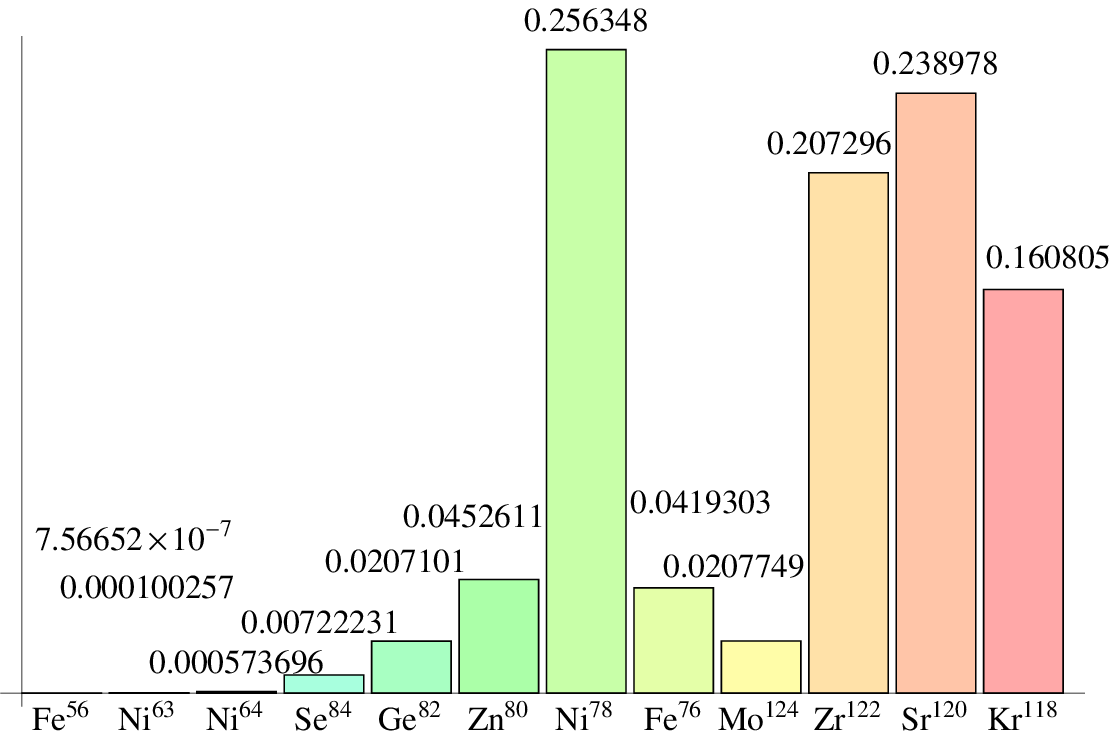}
\includegraphics[width=\columnwidth]{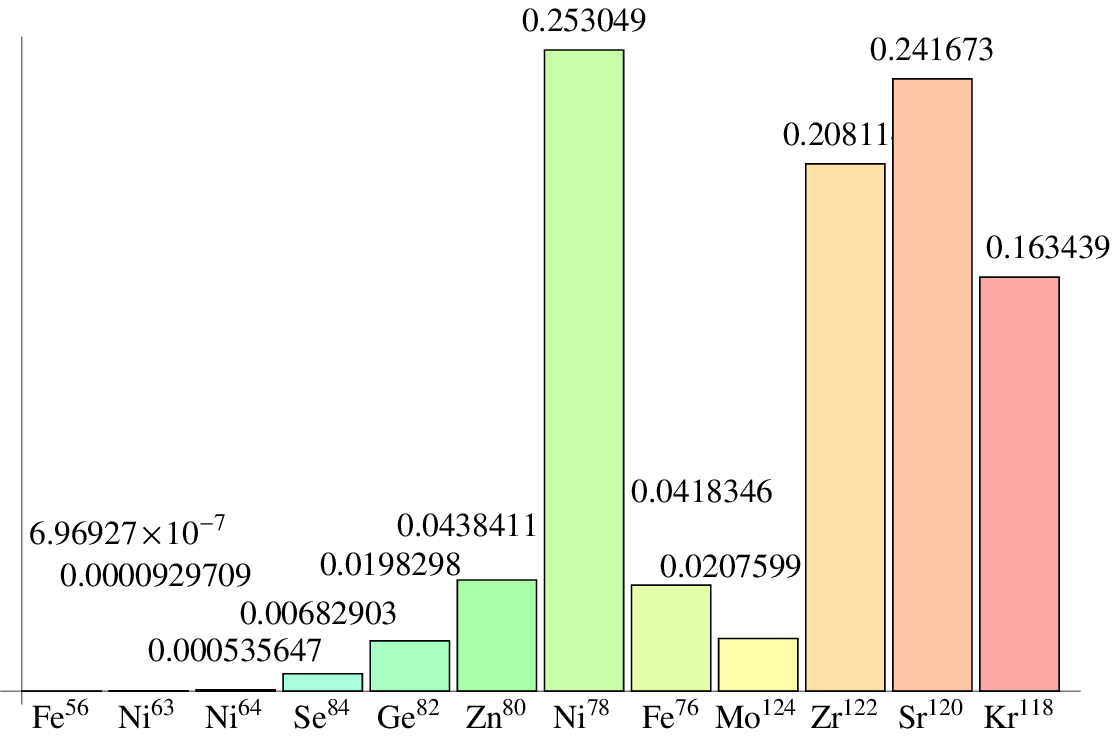}
\caption{Relative abundances of chemical elements in the crust for the two cores analyzed in Table \ref{tab:comparison}}\label{fig:abbondanze}
\end{figure}

In the case of the BPS EoS, the average nuclear composition in the outer crust, namely the average charge to mass ratio of nuclei $Z/A$, is obtained by calculating the contribution of each nuclear composition present to the mass of the crust. We exemplified the analysis for two different cores: $M_{\rm core}=2.56 M_\odot$, $R_{\rm core}=12.79$ km; $M_{\rm core}=1.35 M_\odot$, $R_{\rm core}=11.76$ km. 
The relative abundance of each nuclide within the crust of the star can be obtained as
\begin{equation}\label{eq:thick}
{\rm R.A.}= \frac{1}{M^{\rm BPS}_{\rm crust}}\int_{\Delta r} 4 \pi r^2 {\cal E} dr\, ,
\end{equation}
where the integration is carried out in the layer of thickness $\Delta r$ where the particular nuclide is present; see~\ref{tab:comparison} and Fig.~\ref{fig:abbondanze}. Our results are in agreement with the analysis on the neutron star crust composition obtained in \cite{Goriely11,Pearson11}. In both cases we obtain as average nuclear composition $^{105}_{35}$Br. The corresponding crusts with fixed nuclear composition $^{105}_{35}$Br for the two chosen cores are calculated neglecting Coulomb interactions (i.e. using the first EoS). The mass and the thickness of these crusts with fixed $^{105}_{35}$Br are different with respect to the ones obtained using the full BPS EoS, leading to such average nuclear composition. For the two selected examples we obtain that the mass and the thickness of the crust with average $^{105}_{35}$Br are, respectively, $18\%$ larger and $5\%$ smaller with respect to the ones obtained with the corresponding BPS EoS. This result shows how small microscopic effects due to the Coulomb interaction in the crust of the neutron star leads to quantitative not negligible effects on the macroscopic structure of the configuration.
\begin{table*}
\begin{center}
\begin{ruledtabular}
\begin{tabular}{c c c c c c c}

\multicolumn {7}{c}{Equilibrium Nuclei Below Neutron Drip}\\
    \hline
    Nucleus  & $Z$    & $\rho_{max}$(g cm$^{-3}$) & $\Delta$ $R_1$ (km)& R.A.$1(\%)$            &$\Delta$ $R_2$ (km)& R.A.$2(\%)$ \\
		\hline
		$^{56}$Fe  & $26$ & $8.1 \times 10^6$        & $0.0165$           & $7.56652\times10^{-7}$ & $0.0064$          & $6.96927\times10^{-7}$\\
		
		$^{62}$Ni  & $28$ & $2.7 \times10^8$         & $0.0310$           & $0.00010$              & $0.0121$          & $0.00009$\\
		
		$^{64}$Ni  & $28$ & $1.2 \times10^9$         & $0.0364$           & $0.00057$              & $0.0141$          & $0.00054$\\
		
		$^{84}$Se  & $34$ & $8.2 \times10^9$         & $0.0046$           & $0.00722$              & $0.0017$          & $0.00683$\\
		
		$^{82}$Ge  & $32$ & $2.2 \times10^{10}$      & $0.0100$           & $0.02071$              & $0.0039$          & $0.01983$\\
		
		$^{80}$Zn  & $38$ & $4.8 \times10^{10}$      & $0.1085$           & $0.04521$              & $0.0416$          & $0.04384$\\
		
		$^{78}$Ni  & $28$ & $1.6 \times10^{11}$      & $0.0531$           & $0.25635$              & $0.0203$          & $0.25305$\\
		
		$^{76}$Fe  & $26$ & $1.8 \times10^{11}$      & $0.0569$           & $0.04193$              & $0.0215$          & $0.04183$\\
		
		$^{124}$Mo & $42$ & $1.9 \times10^{11}$      & $0.0715$           & $0.02078$              & $0.0268$          & $0.02076$\\
		
		$^{122}$Zr & $40$ & $2.7 \times10^{11}$      & $0.0341$           & $0.20730$              & $0.0127$          & $0.20811$\\
		
		$^{120}$Sr & $38$ & $3.7 \times10^{11}$      & $0.0389$           & $0.23898$              & $0.0145$          & $0.24167$\\
		
		$^{118}$Kr & $36$ & $4.3 \times10^{11}$      & $0.0101$           & $0.16081$              & $0.0038$          & $0.16344$\\
		
\end{tabular}
\end{ruledtabular}
\caption{$\rho_{max}$ is the maximum density at which the nuclide is present;$\Delta$ $R_1$, $\Delta$ $R_2$ and R.A.$1(\%)$, R.A.$2(\%)$ are rispectively the thickness of the layer where a given nuclide is present and their relative abundances in the outer crust for two different cases: $M_{\rm core}=2.56 M_\odot$, $R_{\rm core}=12.79$ km; $M_{\rm core}=1.35 M_\odot$, $R_{\rm core}=11.76$ km.}\label{tab:comparison}
\end{center}
\end{table*}

%

\section{Observational constraints on the mass-radius relation}\label{sec:4}

It has been recently pointed out that the most up-to-date stringent constraints to the mass-radius relation of neutron stars are provided by the largest mass, the largest radius, the highest rotational frequency, and the maximum surface gravity, observed for pulsars \cite{trumper2011}.

\begin{figure}[h]
\includegraphics[width=\columnwidth]{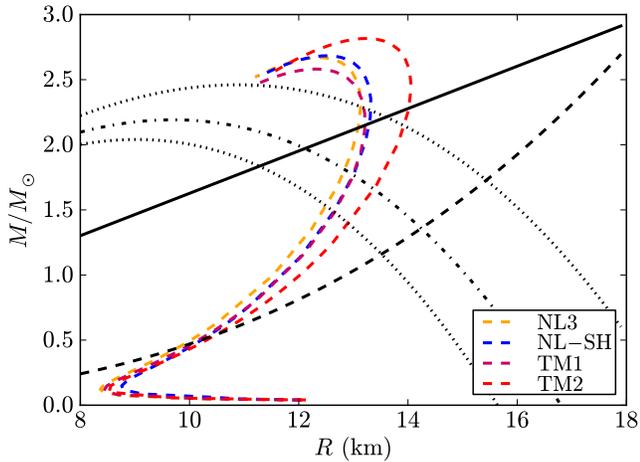}\\
\caption{Constraints on the mass-radius relation given by J.~E.~Tr\"umper in \cite{trumper2011} and the theoretical mass-radius relation presented in this article in Fig.~\ref{fig:star}. The solid line is the upper limit of the surface gravity of XTE J1814-338, the dotted-dashed curve corresponds to the lower limit to the radius of RX J1856-3754, the dashed line is the constraint imposed by the fastest spinning pulsar PSR J1748-2246ad, and the dotted curves are the 90$\%$ confidence level contours of constant $R_\infty$ of the neutron star in the low-mass X-ray binary X7. Any mass-radius relation should pass through the area delimited by the solid, the dashed and the dotted lines and, in addition, it must have a maximum mass larger than the mass of PSR J1614-2230, $M=1.97 \pm 0.04 M_\odot$.}\label{fig:MRconstraints}
\end{figure}

So far, the highest neutron star mass measured with a high level of experimental confidence is the mass of the 3.15 millisecond pulsar PSR J1614-2230, $M=1.97 \pm 0.04 M_\odot$, obtained from the Shapiro time delay and the Keplerian orbital parameters of the binary system \cite{demorest2010}. The fitting of the thermonuclear burst oscillation light curves from the accreting millisecond pulsar XTE J1814-338 weakly constrain the mass-radius relation imposing an upper limit to the surface gravity of the neutron star, $G M/(c^2 R) < 0.24$ \cite{bhattacharyya05}. A lower limit of the radius of RX J1856-3754, as seen by an observer at infinity $R_\infty = R [1-2GM/(c^2 R)]^{-1/2} > 16.8$ km, has been obtained from the fit of the optical and X-ray spectra of the source \cite{trumper04}; it gives the constraint $2G M/c^2 >R-R^3/(R^{\rm min}_\infty)^2$, being $R^{\rm min}_\infty=16.8$ km. Assuming a neutron star of $M=1.4M_\odot$ to fit the Chandra data of the low-mass X-ray binary X7, it turns out that the radius of the star satisfies $R=14.5^{+1.8}_{-1.6}$ km, at 90$\%$ confidence level, corresponding to $R_\infty = [15.64,18.86]$ km, respectively (see \cite{heinke06} for details). The maximum rotation rate of a neutron star taking into account both the effects of general relativity and deformations has been found to be $\nu_{\rm max} = 1045 (M/M_\odot)^{1/2}(10\,{\rm km}/R)^{3/2}$ Hz, largely independent of the equation of state \cite{lattimer04}. The fastest observed pulsar is PSR J1748-2246ad with a rotation frequency of 716 Hz \cite{hessels06}, which results in the constraint $M \geq 0.47 (R/10\,{\rm km})^3 M_\odot$. In Fig.~\ref{fig:MRconstraints} we show all these constraints and the mass-radius relation presented in this article. 

As discussed by J.~E.~Tr\"umper in \cite{trumper2011}, the above constraints strongly favor stiff equations of state which provide high maximum masses for neutron stars. In addition, putting all of them together, the radius of a canonical neutron star of mass $M=1.4 M_\odot$ is highly constrained to the range $R \gtrsim 12$ km disfavoring, at the same time, the strange quark hypothesis for these specific objects. It is clear from Fig.~\ref{fig:MRconstraints} that the mass-radius relation presented here is consistent with all the observation constraints, for all the nuclear parametrizations of Table \ref{tab:models}. We present in Table \ref{tab:MRprediction}, the radii predicted by our mass-radius relation for a canonical neutron star of $M=1.4 M_\odot$ as well as for the millisecond pulsar PSR J1614-2230, $M=1.97 \pm 0.04 M_\odot$.

\begin{table}[h!]
\begin{ruledtabular}
\begin{tabular}{c c c c c}
$M (M_\odot)$ & $R_{\rm NL3}$	& $R_{\rm NL-SH}$ & $R_{\rm TM1}$ & $R_{\rm TM2}$ \\
\hline
1.40 & 12.31 & 12.47 & 12.53 & 12.93 \\

1.93 & 12.96 & 13.14 & 13.13 & 13.73 \\

2.01 & 13.02 & 13.20 & 13.17 & 13.82

\end{tabular}
\end{ruledtabular}
\caption{Radii (in km) predicted by the nuclear parametrizations NL3, NL-Sh, TM1 and TM2 of Table \ref{tab:models}, for a canonical neutron star of $M=1.4 M_\odot$ and for the millisecond pulsar PSR J1614-2230, $M=1.97 \pm 0.04 M_\odot$.}\label{tab:MRprediction}
\end{table}

\section{Comparison with the traditional TOV treatment}\label{sec:5}

In the traditional TOV treatment local charge neutrality as well as the continuity of the pressure and the density in the core-crust transition are assumed. This leads to explicit violation of the constancy of the Klein potentials throughout the configuration (see e.g.~\cite{PLB2011}). In such a case there is a smooth transition from the core to the crust without any density discontinuity and therefore the density at the edge of the crust is $\sim \rho_{\rm nuc}\sim 2.7\times 10^{14}$ g/cm$^3$. The so-called inner crust in those configurations extends in the range of densities $\rho_{\rm drip}\lesssim \rho \lesssim \rho_{\rm nuc}$ while, at densities $\rho\lesssim \rho_{\rm drip}$, there is the so-called outer crust.

\begin{figure}[h]
\includegraphics[width=\columnwidth]{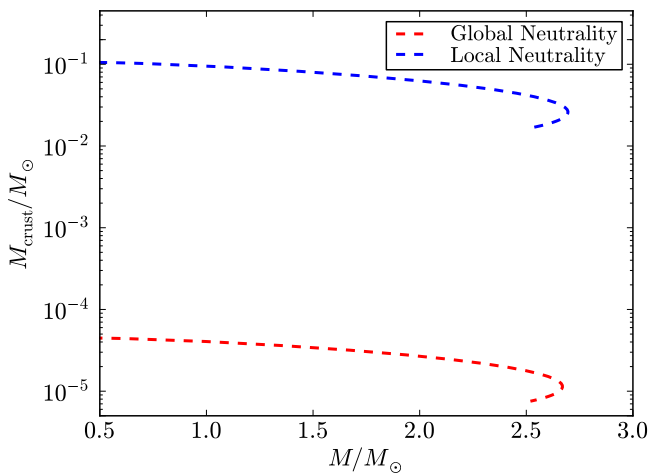}
\caption{Mass of the crust given by the traditional locally neutral Tolman-Oppenheimer-Volkoff treatment and by the new globally neutral equilibrium configurations presented in this article. We use here the NL3 nuclear model, see Table \ref{tab:models}.}\label{fig:innernotinner1_Mcr}
\end{figure}

\begin{figure}[h]
\includegraphics[width=\columnwidth]{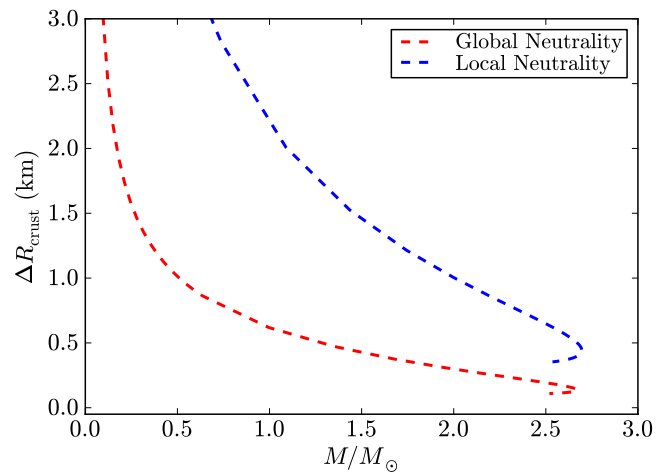}
\caption{Thickness of the crust given by the traditional locally neutral Tolman-Oppenheimer-Volkoff treatment and by the new globally neutral equilibrium configurations presented in this article. We use here the NL3 nuclear model, see Table \ref{tab:models}.}\label{fig:innernotinner1_Rcr}
\end{figure}

In Figs.~\ref{fig:innernotinner1_Mcr} and \ref{fig:innernotinner1_Rcr} we compare and contrast the mass and the thickness of the crust as obtained from the traditional TOV treatment with the new configurations presented here, discussed previously in Secs.~\ref{sec:2} and \ref{sec:3} characterized by $\rho_{\rm crust}=\rho_{\rm drip}$.

\begin{figure}[h]
\includegraphics[width=\columnwidth]{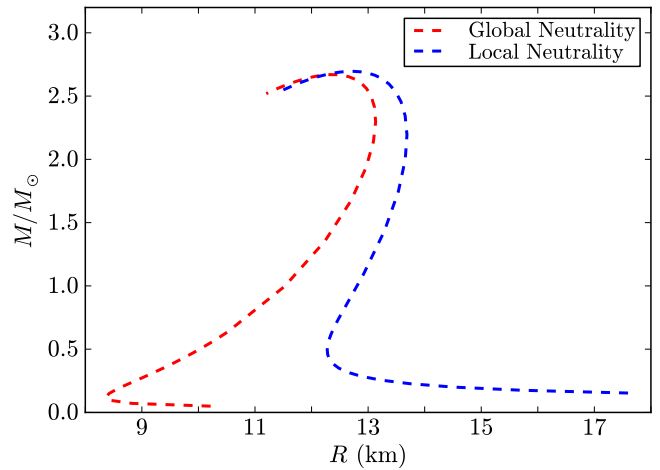}
\caption{Mass-Radius relation obtained with the traditional locally neutral TOV treatment and with the new globally neutral equilibrium configurations presented here. We use here the NL3 nuclear model, see Table \ref{tab:models}.}\label{fig:innernotinner2}
\end{figure}

The markedly differences both in mass and thickness of the crusts (see Figs.~\ref{fig:innernotinner1_Mcr} and \ref{fig:innernotinner1_Rcr}) obtained from the traditional Tolman-Oppenheimer-Volkoff approach and the new equilibrium configurations presented here, leads to a very different mass-radius relations which we compare and contrast in Fig.~\ref{fig:innernotinner2}.

\section{Concluding Remarks}\label{sec:6}

We have formulated the equations of equilibrium of neutron stars based on our recent works \cite{NPA2011} and \cite{PRC2011,wd2011,PLB2011}. The strong, weak, electromagnetic, and gravitational interactions are taken into due account within the framework of general relativity. In particular, the strong interactions between nucleons is described by the exchange of the $\sigma$, $\omega$, and $\rho$ mesons. The equilibrium conditions are given by the set of Einstein-Maxwell-Thomas-Fermi equations and by the constancy of the general relativistic Fermi energies of particles, the Klein potentials, throughout the configuration. 

We have solved these equilibrium equations numerically, in the case of zero temperatures, for the nuclear parameter sets NL3 \cite{lalazissis97}, NL-SH \cite{sharma93}, TM1 \cite{sugahara94}, and TM2 \cite{hirata95}; see Table \ref{tab:models} for details. 

A new structure of the star is found: the positively charged core at supranuclear densities is surrounded by an electronic distribution of thickness $\gtrsim \hbar/(m_e c)\sim 10^2\hbar/(m_\pi c)$ of opposite charge and, at lower densities, a neutral ordinary crust. 

In the core interior the Coulomb potential well is $\sim m_\pi c^2/e$ and correspondingly the electric field is $\sim (m_p/m_{\rm Planck}) (m_\pi/m_e)^2 E_c \sim 10^{-14} E_c$. Due to the equilibrium condition given by the constancy of the Klein potentials, there is a discontinuity in the density at the transition from the core to the crust, and correspondingly an overcritical electric field $\sim (m_\pi/m_e)^2 E_c$ develops in the boundary interface; see Fig.~\ref{fig:electric_field1}--\ref{fig:electric_field2}. 

The continuity of the Klein potentials at the core-crust boundary interface leads to a decreasing of the electron chemical potential and density, until values $\mu^{\rm crust}_e < \mu^{\rm core}_e$ and $\rho_{\rm crust}<\rho_{\rm core}$ at the edge of the crust, where global charge neutrality is achieved. For each central density, an entire family of core-crust interface boundaries and, correspondingly, an entire family of crusts with different mass and thickness, exist. The larger $\rho_{\rm crust}$, the smaller the thickness of the interface, the peak of the electric field, and the larger the mass and the thickness of the crust. The configuration with $\rho_{\rm crust}=\rho_{\rm drip}\sim 4.3\times 10^{11}$ g/cm$^3$ separates neutron stars with and without inner crust. The neutron stars with $\rho_{\rm crust}>\rho_{\rm drip}$ deserve a further analysis in order to account for the reduction of the nuclear tension at the core-crust transition due to the presence of dripped neutrons from the nuclei in the crust. 

All the above new features lead to crusts with masses and thickness smaller than the ones obtained from the traditional TOV treatment, and we have shown specifically neutron stars with $\rho_{\rm crust}=\rho_{\rm drip}$; see Figs.~\ref{fig:innernotinner1_Mcr}--\ref{fig:innernotinner1_Rcr}. The mass-radius relation obtained in this case have been compared and contrasted with the one obtained from the locally neutral TOV approach; see Fig.~\ref{fig:innernotinner2}. We have shown that our mass-radius relation is in line with observations, based on the recent work by J.~E.~Tr\"umper \cite{trumper2011}; see Fig.~\ref{fig:MRconstraints} for details.

The electromagnetic structure of the neutron star presented here is of clear astrophysical relevance. The process of gravitational collapse of a core endowed with electromagnetic structure leads to signatures and energetics markedly different from the ones of a core endowed uniquely of gravitational interactions; see e.g.~\cite{vitagliano03a,vitagliano03b,ruffinireview08,physrep}.

It is clear that the release of gravitational energy in the process of gravitational collapse of the core, following the classic work of Gamow and Schoenberg (see \cite{gamow41,arnettbook}), is carried away by neutrinos. The additional nuclear and electromagnetic energy $\sim 10^{51}$ erg of the collapsing core introduced in this article are expected to be carried away by electron-positron plasma created in the overcritical electromagnetic field in the collapsing core.



\end{document}